\documentclass[twocolumn]{emulateapj}
\submitted{Science with Parkes @ 50 Years Young, 31 Oct. -- 4 Nov., 2011}

\shorttitle{The First Observations with the GRT at Parkes}
\shortauthors{R. M. Price}

\begin{document}

\title{The First Observations with the GRT at Parkes}

\author{R. M. Price}

\affil{University of New Mexico, Physics and Astronomy, Albuquerque NM, USA}

\email{priceabq@aol.com}

\begin{abstract}
In this contribution Marcus Price gives a first hand account of some of the very first scientific obervations undertaken with the ``Giant Radio Telescope'' at Parkes. It was clearly a very exciting time of discovery, enabled by superb engineering. 
\end{abstract}
\keywords{history and philosophy of astronomy}

\section{Introduction}

The Parkes 210 foot dish was nearing completion on the opening day in October of 1961.  Mechanically, the dish and the tower were complete.  The dish could be operated manually in alt-az mode.  Within a month, the control desk (all analogue) and the servo system were completed and undergoing testing and final tune-ups. The commissioning receivers were installed and tested.  They were a 21 cm crystal receiver and a 75 cm crystal mixer receiver, connected to an on-axis crossed dual dipole feed.

Aside from zenith drifts, the first observations with the GRT were drift scans of the radio source Fornax A (NGC 1316), using only the alt-az system.  Soon afterwards, the servo system and Master Equatorial were operational and formal scientific observations commenced.

\section{Arrival at Radiophysics}

I arrived at the Radiophysics Laboratory as a Fulbright Scholar at the end of August 1961.  I had a B.S. Degree in Physics, and some research experience in radio propagation gained by a brief period of employment at the Central Radio Propagation Laboratory of the U.S. National Bureau of Standards.   I had arranged to work with Joe Pawsey at Radiophysics.  These plans quickly changed with Pawsey’s sudden illness.  I did start on my my initial project, a comparison the southern results of the Mills Cross with the 2C catalogue.  This project came quickly to a close (for several reasons!), and John Bolton took over my direction.

John Bolton believed in a ``hands on'' approach.  I was sent to the electronics group, starting in the transformer shop, and quickly working up to testing microwave switches, transformers, and loads.  My ‘spare time’ was spent in courses in EE and Physics at The University of Sydney (easy since the Radiophysics Laboratory was on the University grounds.) .  The excellent library at RP allowed me to become familiar with the history of radio astronomy and start to follow the current journals.

In mid October, Bolton told me to move from Sydney to the GRT at Parkes.  In my old Land Rover and the roads of the time this turned into a two day trip.  I arrived and checked into the Star Hotel (the quarters were full with CSIRO staff).  I was assigned general duties at the dish--mostly cleaning up construction debris in the tower and  washing the windows in preparation for the opening day.

\section{Opening Day}

Opening day arrived with 60 mph winds.  The dish could not be dipped to the horizon  to salute the Governor General as had been planned. But, the skill of a Canadian, Norm Broten, in climbing to the aerial cabin in the gale allowed the Australian flag to be proudly flown above the dish.

The opening was highly structured.  Depending on the color of your invitation name tag, you were assigned to one of three groups:  VIPs, special visitors, or general public.  The VIPs got a tour of the tower, a seat at the ceremony, and lunch, special visitors got the ceremony and lunch, and the public got to watch.  Because of the coolness between RP and Sydney University School of Physics, all of the University visitors were in the general public category.  However, this did not stop Harry Messel (Head of Physics at Sydney U.) from joining the VIPs and holding forth at every opportunity as if he had personally designed and funded the telescope.

With the opening completed, the RP staff could now get down to business.  I moved out to the quarters where I was to spend much of the next four years.  I split my time between the control system engineer, Jack Rothwell, and the control desk electronic technician, Klaus Kalweiht (sp?).  In this way I quickly leaned the control and servo systems, down to knowing how to balance the valve servo amplifiers and, perhaps more importantly,  where all of the fuses were.  All of this was before computers and the digital age.  The control system was one big analog computer with the master equatorial unit at its heart.

On the receiver side, I helped install the initial crystal mixer receiver and feed systems that operated at 21 and 75 cm.  Being the youngest staff member, it fell to me to make trips at all hours to the aerial cabin to replenish the liquid air used for the cold loads on the receivers.  The receiver output was recorded on two pen chart recorders. Time and position information was recorded on the charts by hand.   Information was later read by eye and transferred to large pieces of graph paper to draw contours to make brightness temperature maps.

\section{The First Scientific Observations}

After several weeks, Bolton wanted to get started with observations even though the control desk was not fully functional yet.  He decided to look at Fornax A.  This necessitated calculating the appropriate alt-az coordinates to start a number of drift scans at a separation of about 7 arcminutes (the half power beam [FWHM] of the GRT at 21 cm was approximately 14 arcminutes).  Again,  as the junior member of the team, it fell to me to calculate the scan start coordinates.  This was before the age of hand held calculators, so I used an old Marchant calculating machine.  (The kind that goes chuga-chuga-chuga zwoop, when dividing.)  A long and tedious  set of calculations in spherical trigonometry filled the morning hours.

We then ‘drove’ the dish to the first scan start point.  We waited, and soon the 75 cm signal increased, followed by the 21 cm signal in a few minutes.  Each showed a nice beam shape.  We moved to the second scan.  We waited, and soon we got a larger 75 cm signal, and a double 21 cm profile!  (Recall, that at this time it was not clearly established that powerful radio sources were often doubles.)   We moved again, and again the  unresolved 75 cm signal and the double 21 cm signal.  We moved to next  scan start.  We waited...and waited.  Nothing.  That was the first time I received the Bolton whammy look.  The calculated alt-az coordinates clearly were incorrect.  We moved to the next scan, and the signals returned.  By,  the end of the afternoon, with no more missed scans, we had a fair map of Fornax A at 21 cm and the relative flux densities at 75 and 21 cm.
(Fortunately, for my career in radio astronomy, soon after that, Bolton was using the Marchant and found that it was sometimes unreliable.)

Within days, the control system was fully operational, although many ‘bugs’ remained.   Up until this time  the observing ‘teams’ consisted of whom ever was at Parkes from the Sydney lab, John Bolton, and myself.  Usually this was dependent on what receivers were being installed or worked on.  But for full scale observations, Bolton assigned particular people and groups  to projects.  The groups consisted of several  people, e.g  Frank Gardner and John Whiteoak, mostly with technical backgrounds.  Only a few of the Radiophysics staff had an astronomy background, e.g. Jim Roberts, John Whiteoak, and Don Matthewson.  Others, like Brian Cooper and Dick McGee came from the technical side.  (Since the late 1940's, the staff at Radiophysics had been inventing radio astronomy as they went along--they were among the first generation of radio astronomers,  made up from physicists, engineers and mathematicians.)  Establishing calibration standards was a task we all shared--with John Bolton being the lead.  John Shimmons worked on the pointing and position calibrations.  And finally, several groups conducted a fully sampled  survey of the entire southern sky at 75 cm wavelength, along with the additional information obtained with the 21 cm receiver (thanks to the twin dual dipole feed).

At first, observing could only be accomplished with the help of the servo engineer in the control room.  But, slowly the system became reliable enough for the Radiophysics people to operate. In the interim, a notice appeared on the bulletin board-- ‘The dish could be operated only if  John Bolton or Marc Price was in the control room.’   I became the regular second half driver for the first six months or so.  This for two reasons,  I was the most junior person around, and I knew all of the idiosyncrasies of the control system.

Being at Parkes was the best education a graduate student could ask for.  In addition to the hands-on experience of observing with the world’s best radio telescope, every tea break or meal was a tutorial.  Discussions always centered on radio astronomy and the latest results from Parkes or elsewhere.  Often visitors from observatories all over the world were present and added to the knowledge shared at these events.

\section{Major Discoveries of the First Year}

There were numerous discoveries in the first year of observing.  Of course, the surveys were finding numerous unknown sources.  One of the most important was 1934-63,  one of the sources discovered in the -60 to the south celestial pole survey (SCP) being carried out by Doug Milne and myself.  Ken Kellermann gives details of that source in his paper in this volume.

Another major discovery in the first year was the Faraday Rotation of the polarization in the emission from NGC5128, Centaurus A.  Brian Cooper and I were assigned Cen A by Bolton, and we were just beginning to survey this large source (over 9 degrees of declination on the Southern sky).  On that same weekend, a former Radiophysics scientist, Ron Bracewell arrived from Stanford U. to visit the telescope.  Bolton was not around and it was not clear if Bracewell was authorized to observe with the telescope.  But, Brian Cooper erred on the side of graciousness and let Bracewell observe.  The new feed rotator had just been installed, and he was able to survey the inner source of Cen A with the 10 cm receiver,  The asymmetric polarization of the double source stood out clearly with the resolution and sensitivity of the Parkes telescope.  By breakfast time, Bracewell had a map of the central source.  

Since it was the Easter weekend.  The telescope was not observing on the next day, Easter Sunday.  I didn't want  to see this new discovery go untested.  I obtained the assistance of the Site Manager, George Day, to be the second man in the control room.  I then changed the feed and receiver to 21 cm.  Although the resolution was not as good, it was clear that the polarization was there.  But, there was a problem,  I thought Bracewell had recorded the position of the feed rotator incorrectly.  The polarization angle was just 90 degrees different from what he had recorded.
.
The next few days were filled with confusion and astonishment.  Bolton returned to find the Bracewell result, and my report of the position angle ‘error.’  Repeat measurement quickly showed that Bracewell and I were both correct in our measurements.  The only answer seemed to be Faraday Rotation.   But, both observations were carried out by non-CSIRO people and under questionable circumstances.  (To my knowledge it has never been determined if ‘Taffy’ Bowen, head of Radiophysics actually granted Bracewell observing rights, or if he only intended that Bracewell ‘observe’ the operations at the telescope.)  In the case of the Easter observations I had made, I later learned that the telescope had been officially closed for the Easter weekend!

Brian Cooper and I then did a comprehensive study of the Cen A central source at half a dozen frequencies to show without doubt that it was the Faraday effect we were seeing.  Of course, we also studied the polarization properties of the entire source over the following months for later publication.

Gardner and Whiteoak were quickly assigned to study polarization and Faraday Rotation in radio sources.  Within weeks they submitted a paper on their initial studies to Physical Review Letters.  It was published before either Bracewell’s Polarization paper  or Cooper and Price’s Faraday Rotation discovery paper.  The power and flexibility of the Parkes dish was confirmed.

\section{My Thesis Project}

I would be remiss and uncharacteristically humble not to mention another almost first at the Parkes Radio Observatory.  In spite of the numerous other projects I joined in at Parkes, for my Ph.D. research at Mount Stromlo (Australian National University) I chose to study the radio emission properties of the disk and halo of our Milky Way.  This, of course, fit right in with the interests of my co-supervisors John Bolton and B. J. Bok.

For this study, I designed, built, and calibrated a large standard gain horn to measure the  absolute sky brightness at 75 cm (408 MHz).    In addition to building the horn, I also had to design and build a coaxial switch and a matched load to go with the standard crystal mixer receiver I used.  Using the SCP as a reference field in the sky, I could compare the 75 cm survey Milne and I had done of that region, with the horn result.  This allowed me to determine the  minimum temperature in the southern sky as measured with the Parkes dish to be about 10 K.  My measurements came a few months after the initial observations of Penzias and Wilson and before they were certain of their 3 K measurement.  It was obvious a few months later when they published their results, that 3 K of my 10 K minimum was the CBR.  Of course, because of the non-thermal spectrum of the galaxy and radio galaxies at 75 cm, there was far too much foreground (galaxy) and background (unresolved distant radio sources) to separate out the CBR in my measurement.  But, it still felt good to measure `my' absolute temperature with apparatus that I had built almost entirely by myself, whereas it took the technical might of Bell Labs to facilitate the Penzias/Wilson measurement.

\section{Epilogue}

In conclusion. During the early 1960's I had the great pleasure and privilege of working with a number of the world's best radio astronomers and microwave engineers and to use the worlds best single dish radio telescope.  Added to that, the activities of the Mount Stromlo Observatory under the direction of Bart J. Bok, led to a tremendous educational opportunity for a young graduate student.  My thanks and gratitude to all those pioneers  of radiophysics and radio astronomy in Australia in the 1950's and 60's.

\end{document}